# The Nature of Complexity in the Biology of the Engineered Nanoscale Using Categorization as a Tool for Intelligent Development

**Kenneth A. Dawson**

Centre for BioNano Interactions, University College Dublin, Ireland

Email: kenneth.a.dawson@cbni.ucd.ie

ORCID iD: 0000-0002-0568-6588

**Abstract**

Throughout the evolution of biological species on Earth, cells and organs have developed many complex structures and processes to ensure their interactions with individual chemical molecules (small and macromolecular) and nanoscale objects result in no harm. These evolutionary mechanisms complicate our attempts to use modern nanoscale science to develop effective and efficient treatments for disease or other biological dysfunctions. Here we describe the complexity of biology on the nanoscale and the implications for the success of recently-discovered nanoscience, which has resulted in an almost infinite number of potential nanomaterials of unknown efficacy. We discuss how tools to categorize nanomaterials on the basis of structure, properties and interactions can provide insights on promising directions.

Those seeking to discover and understand unique nanoscale interactions in biology that go beyond common biological paradigms of receptor-ligand and recognition-induced signalling are, in a sense, addressing a key question: *Whether the discipline of nanoscale biology is a durably distinct science that will grow in decades to come to have its own fundamental knowledge that enables great advances in science or will it slowly but solely merge and become part of the rich tapestry of the scientific journey, and elevate all biology?* Possibly the best outcome for those excited by the field is both, but the real question remains: *What can the engineered nanomaterial do that nothing else and no competing technology achieve?*

For those more focused on the challenges of ensuring safety in nanotechnology, the scientific questions are similar, but with a different focus. Succinctly put, can increasingly complex engineered nanomaterials exhibit unforeseen outcomes because they interact with the environment in a different way?

After a decade and a half of effort [1-11], it is increasingly clear that biological interactions on the nanoscale are simply different. The fact that they are truly and profoundly different, initiated differently, and processed differently carries with it the message for the future. Here we are not seeking to convince an audience that our mastery of those interactions is sufficient to achieve all that is hoped from the arena. We instead wish to share a view, gathered from those years, about where we are headed in this whole question, with confidence that the future, some of it closer than we expect, will confirm many of the points.

At the time of writing, the key salient points seem to be that biology has engineered into itself the capacity to deal with, make use of, and protect itself exquisitely from nanoscale objects. For those unfamiliar with the field, perhaps the arena of 'exosomes' (small-cell-derived vesicles of some tens of nanometers) will illustrate how biology already has a parallel and effective system of communications and control implemented by explicitly nanoscale (albeit endogenous) structures [12, 13].

That system is all quite a different paradigm from that for small molecule chemicals which can override all the rules of biology by simply 'dissolving' in the living organisms and leaving those organs responsible (such as the liver) to clean up the consequences. Moving into the nanoscale we enter a new realm in which trafficking, immune reactions, and specific protective gatekeeping mechanisms of all kinds come into play. The complexity of those mechanisms at first were a surprise to the scientific community.

It is important to understand the distinction a cell makes between a simple protein (e. g. albumin, around 7 nm) that is certainly at nanoscale and the 'complex nanoscale objects' in which, for instance, a tapestry of numerous proteins assemble at the surface or possess irregular structural features on that scale of tens of nanometers. We now surmise that some of these naturally occurring complex (exosomal) nanostructures can act to regulate the immune system and be exploited by tumours to silence our defences, while others contribute to multiple layers of defence [14-17]. The enormous power of nanoscale biology, however, should not be conflated with the idea that it will lead (effortlessly and phenomenologically) to new cures, nor (accidentally) to new hazards. We will return to that point below.

We are confident that the next few years will progressively unveil some of these phenomena. Much of the focus will of course be, as it started, in medicine design and efficacy, and a new-found realism of the type of precision and intelligence of design required to tap meaningfully into these processes and use them for benefit. There will always be allied to this fundamental thinking the question of whether such complex objects could induce new forms of harm or constitute new hazards. It is hardly surprising that nanoparticles made of toxic materials are also toxic [18], but whether long lived and slowly degradable new nanostructures of great complexity (of surface, shape, mechanical properties) could present new risks remains to be seen. There may be some buried hazards, but we need some basic hypotheses to predict what we are looking for. The asbestos phenomenon is perhaps a good analogy: a (thankfully) rare but disturbing substance that exhibits a surprising negative biological impact. We should be on the lookout for new, but similar, examples and paradigms in the years to come.

Paradoxically, our understanding has grown to lead us in the opposite direction of 'obvious' outcomes. We are beginning to have some basic understanding allowing us to build hypotheses around which nanostructures could be a medical benefit, and (credibly) possibly even predict possible harm [19-21]. Informing all of this is a growing belief that perhaps some readers will find surprising. The great abundance of complex nanostructures to which biology has had to respond over long ages have been hazardous pathogens or other threats. We are now convinced that coevolution with nanoscale pathogens has led to an astoundingly complex endogenous machinery, operating on this nanoscale, that can deal with complex objects of that size [22]. We see everywhere in our deepening understanding of nanoscale biology the fingerprint of long-fought, hard-won wars of survival in which growing the co-evolution of complex threat was matched by complex biological countermoves [23-25]. At the time of writing this article, the world is confronting a new virus (Covid-19) and while all are aware of the risks involved, we fully expect our natural endogenous defences to finally win, and like so many previous threats, for the virus to recede.

The fundamental point here is that working in the realm of engineered nanoscale biology, we are increasingly aware of the complexity and sophistication of the defences. Those defences are immeasurably more sophisticated on the nanoscale than on the small molecule chemical scale. For those developing medicines, this was at first a considerable disappointment as the major task became to be smarter and more accomplished in precision nanoscale engineering than the organisms and pathogens that have confronted each other throughout the history of living organisms [26-29]. That discovery has turned out to be challenging, but with sufficient humility and a willingness to learn from those arenas, we are now making progress in the direction of new therapies.

On the side of 'new hazard,' the implications were also surprising for many. Far from being a novel and ubiquitous threat, nanoscale engineered entities are quickly transformed in the body to familiar enemies, typically of such mundane levels of threat that it was barely worth raising all but a 'default' response to eliminate them [30-32]. Our focus in the safety arena is now much clearer. We see that hazards, unlike chemicals, if they exist in engineered nanostructures to an exceptional level, will come from the truly surprising, the 'check mate' move that surprises all, that for millions of years were not contemplated by pathogens nor by the cells to defend against. Indeed, the potential threat exists to turn those well-honed defences against ourselves [33-36]. As with the recent Covid-19 threat and with asbestos, it may be the over-reaction of our defences that leads to trouble, and the ill-judged capacity of the defence mechanism to know when to stop the danger signals of cytokines and other components there to protect us. There may indeed be other subtle 'check mates' of the established protective mechanisms.

We are relatively sure, though, that these will not be simple outcomes of simple known mechanisms known from old-style toxicity studies. So therein lays a considerable paradox given the history: that finding a guarding against environmental or human harm of nanostructures may well be an honourable objective for that rare lifelong dedicated professional detective, with high scientific method, rather than the ubiquitous ease of tests. That will certainly be an interesting story for our children to tell, if they even recall the wave of suspicion with which nanostructures were originally met.

For those with a long-term and dedicated purpose to protect against harm, or for those whose objective is to finally tap in the universe of complex nanostructures and exploit them for good, whether it be to reverse the cunning of malign exosome in tumour-immune regulation, or the multitude of diseases as yet barely touched by current therapies, the agendas will remain intermingled.

Without doubt, one of those shared objectives will be to bring some order to the potentially unlimited types of nanostructures we can make or are made by design or accident in the modern world. That will ultimately point towards the human desire to 'categorise' these objects into something we can handle or even contemplate. Of course, the desire to categorize nanostructures is not unique to biology, medicine or related disciplines. Those involved in making new devices will have their own interests. Here in the heart of the search for beneficial biological impacts, the desire to categorise is long and deep. It began decades ago with Chemistry and its desire to learn from previous chemicals of interest, both drugs and toxins, and the belief that shared motifs of chemical identity and structural identity point towards new drugs and new hazards [37-40]. That has to be considered a tremendous success story of categorization. The emergence of automated intelligent searching tools has accelerated, culminating in recent reports that new antibiotics are being prompted simply by categorization and 'intelligent' comparison [41-43].

The challenges for nanostructures are their almost unlimited numbers and complexity. Bounded or limited by no such rules of stoichiometry and bonding, the size, shapes, surfaces, (nano) mechanical, and all other properties suggest no natural limits on the types we can make, or accidentally produce [44-48]. It is indeed, we believe, a profound and fortunate outcome that nanoscale is so well policed and monitored in biology, for in the absence of such we could face a truly unknown and unlimited set of outcomes.

Our modern understanding now focuses much more on biased sets of categories where growing scientific confidence suggest arenas of interest. For instance, complex biologically coated surfaces (either purposefully or by exposure to organisms), as well as complex nanoscale shape in which structural features are visible on the scale of the particle itself, seem to offer platforms within which to categorise [49]. That is leading to a whole new set of concepts and tools about the subcategories within those broad spaces of parameters, and that endeavour will no doubt accelerate. A recent effort to begin the computation categorization of shape [50], and link that to biology is illustrated in Figure 1, and likely points to future progress.

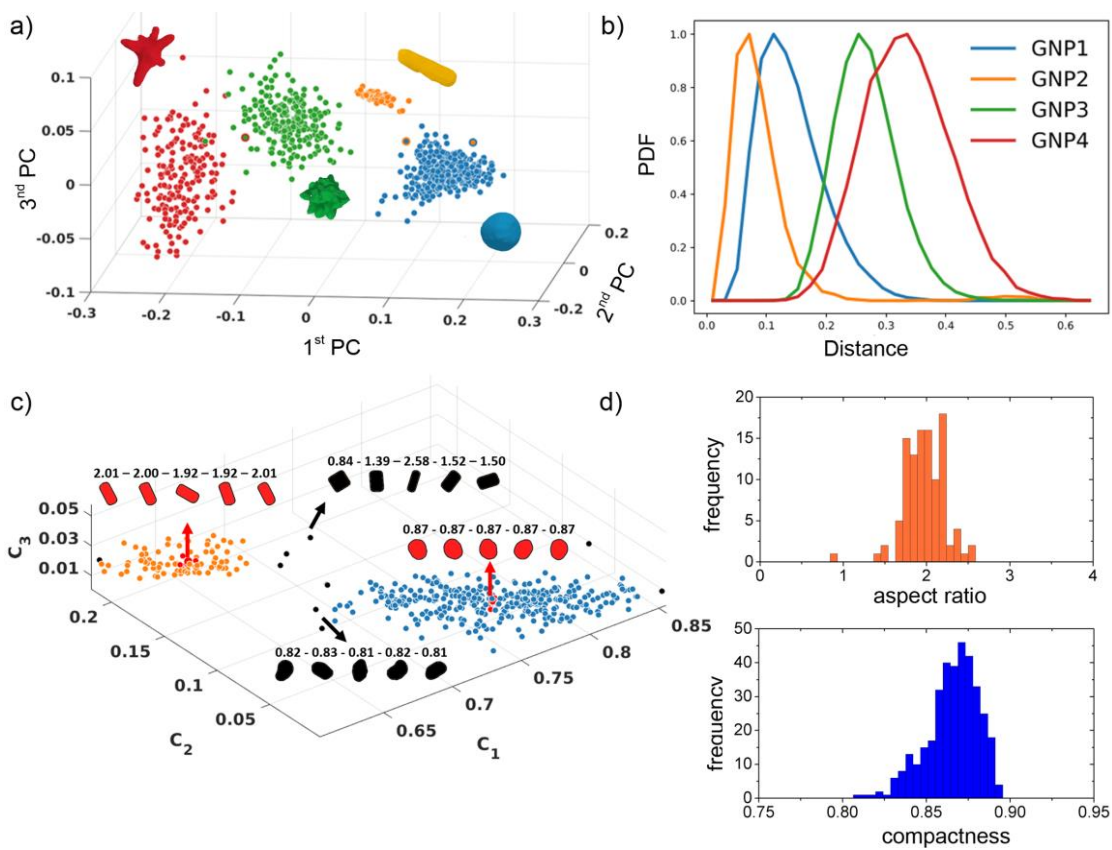

**Figure 1:** *A priori* identification of nanoparticle shape groups from a large contour database of different shapes derived from randomised multiple different syntheses [50] a) Clustering using center of gravity method represented as scatter plots for the first three principal components (PCs). Each dot represents one individual NP. Shape group colors are GNP1 (blue), GNP2 (orange), GNP3 (green), GNP4 (red). The first three PCs with magnitude and phase are represented in the expanded space (variance = 88%); b) Normalized probability distribution function (PDF) of shape variability for each shape group. c) Graphical representation of the contours of 5 particles from the centre (red) and 5 particles from the periphery (black) of the cluster of GNP1 and GNP2. On top of each GNP1 particle, the relative values of compactness (1 for a perfect sphere) and on top of each GNP2 particle the relative values of aspect ratio. d) Distribution of aspect ratio for GNP2 and compactness for GNP1. Figure originally appeared in Boselli L, Lopez H, Zhang W et al (2020). Commun Mater 1, 35 [50]

There are, however, quite different ideas on the horizon. Intelligent and automated analysis of 'big data' may well lead to new insight and *de novo* ways of thinking about categorization [51-53]. The problem is, for the moment, the paucity of good quality data [54-57], in common format, and many efforts to go along the big data direction will be cramped for some years to come, unless that data availability problem can be changed.

The articles of this collection are a worthy reminder that many parts of life await progress with great anticipation. For many in the field, categorizing nanomaterials, with some emphasis on their likely biological (or hazard) outcomes, is a key and living priority.

Rumble discusses that there are many dimensions possible for categorization of nanomaterials, and the context of the categorization effort strongly influences how successful a specific scheme is. For example, as Brown points out one goal of today's nanoscience is to develop innovative materials with properties and functionalities far beyond what is present in existing materials. Thus, categorizing nanomaterials with respect to potential properties and functionalities is a context that supports innovation and creativity. In contrast, Peijnenburg argues that the large number of potential nanomaterials precludes individual testing of each new substance, which makes it important to categorize on the basis of what is potential their most important property – risk to the environment. The papers by Doa, Sargent and Kobe all focus on the fact that most countries today have fairly stringent regulatory regimes with respect to nanomaterials. Consequently, effective categorization for regulatory efficiency must be pursued to ensure that regulation does not stifle innovation and creativity.

Biological complexity is a major challenge for future nanoscience, whose promise for benefits and positive impacts remains bright. The power of today's research community is that the result of the creativity and innovations of the thousands of individual scientists working on the nanoscale is larger than the sum of the individual efforts. The synergy and energy from these efforts will produce results far beyond our imagination. To achieve this promise, we must use all the tools we have to guide us, individually and collectively, in the right directions. As the papers in this collection describe, categorization is a tool for progress, but we must always be aware of the best use of this tool.

**Acknowledgements**

This article is one of a collection of articles about the categorization of nanomaterials, generated by research and workshop discussions under the FutureNanoNeeds project funded by the European Union Seventh Framework Programme (Grant Agreement No 604602).

Author declares that he has no conflict of interest.